\documentclass[aps,prb,preprint]{revtex4-1}
\usepackage{graphicx}
\usepackage{amsmath}
\usepackage{amsfonts}
\usepackage{color}

\begin{document}

\title{Effect of strain and doping on the polar metal phase in LiOsO$_3$}

\author{Awadhesh Narayan}
\email{awadhesh@iisc.ac.in}
\affiliation{Solid State and Structural Chemistry Unit, Indian Institute of Science, Bangalore 560012, India,}
\affiliation{Materials Theory, ETH Zurich, Wolfgang-Pauli-Strasse 27, CH 8093 Zurich, Switzerland.}

\date{\today}

\begin{abstract}
We systematically investigate the effect of strain and doping on the polar metal phase in lithium osmate, LiOsO$_3$, using first-principles calculations. We demonstrate that the polar metal phase in LiOsO$_3$ can be controlled by biaxial strain. Based on density functional calculations, we show that a compressive biaxial strain enhances the stability of the polar $R3c$ phase. On the other hand, a tensile biaxial strain favors the centrosymmetric $R\overline{3}c$ structure. Thus, strain emerges as a promising control parameter over polar metallicity in this material. We uncover a strain-driven quantum phase transition under tensile strain, and highlight intriguing properties that could emerge in the vicinity of this polar to non-polar metal transition. We examine the effect of charge doping on the polar metal phase. By means of electrostatic doping as well as supercell calculations, we find that screening from additional charge carriers, expected to suppress the polar distortions, have only a small effect on them. Rather remarkably, and in contrast to conventional ferroelectrics, the polar metal phase in LiOsO$_3$ remains robust against charge doping up to large doping values.
\end{abstract}

\maketitle

\section{Introduction} 

Ferroelectrics are an important class of insulating materials with a spontaneous polarization, which can be switched by an external electric field~\cite{lines2001principles}. Usually, as temperature is lowered, a material can transition from a paraelectric to a ferroelectric phase, with an accompanying loss of inversion symmetry. In a seminal paper, Anderson and Blount proposed that a similar continuous structural phase transition can occur in metallic systems~\cite{anderson1965symmetry}. Motivated by their proposal, several materials including V$_3$Si~\cite{testardi1975structural,paduani2008martensitic}, Nb$_3$Sn~\cite{testardi1975structural}, and Cd$_2$Re$_2$O$_7$~\cite{sergienko2004metallic,tachibana2010thermal}, were investigated as potential candidates. A clear example of such a ``polar metal'' was demonstrated by Shi \textit{et al.} in a high-pressure-synthesized material lithium osmate, LiOsO$_3$~\cite{shi2013ferroelectric}.

Motivated by these discoveries, there have been several theoretical suggestions, employing first-principles calculations, to design systems exhibiting a polar metal phase. These include cation ordered ruthenate oxide~\cite{puggioni2014designing}, as well as layered perovskites~\cite{filippetti2016prediction,fang2019electric}. Furthermore, it has been suggested that polar metals could be used for potential applications, such as creating a Mott multiferroic~\cite{puggioni2015design} and as electrodes in ferroelectric nanocapacitors~\cite{puggioni2018polar}. In addition there have been notable advances on the experimental front. ``Geometric design'' of such polar metal phases in thin film structures has been realized recently~\cite{kim2016polar}. Guided by first-principles calculations, Kim \textit{et al.} found a polar metal phase in perovskite nickelates~\cite{kim2016polar}. Very recently, Cao \textit{et al.} have demonstrated a two-dimensional variant of the polar metal phase in tri-color superlattices~\cite{cao2018artificial}.

\begin{figure}
\includegraphics[scale=0.6]{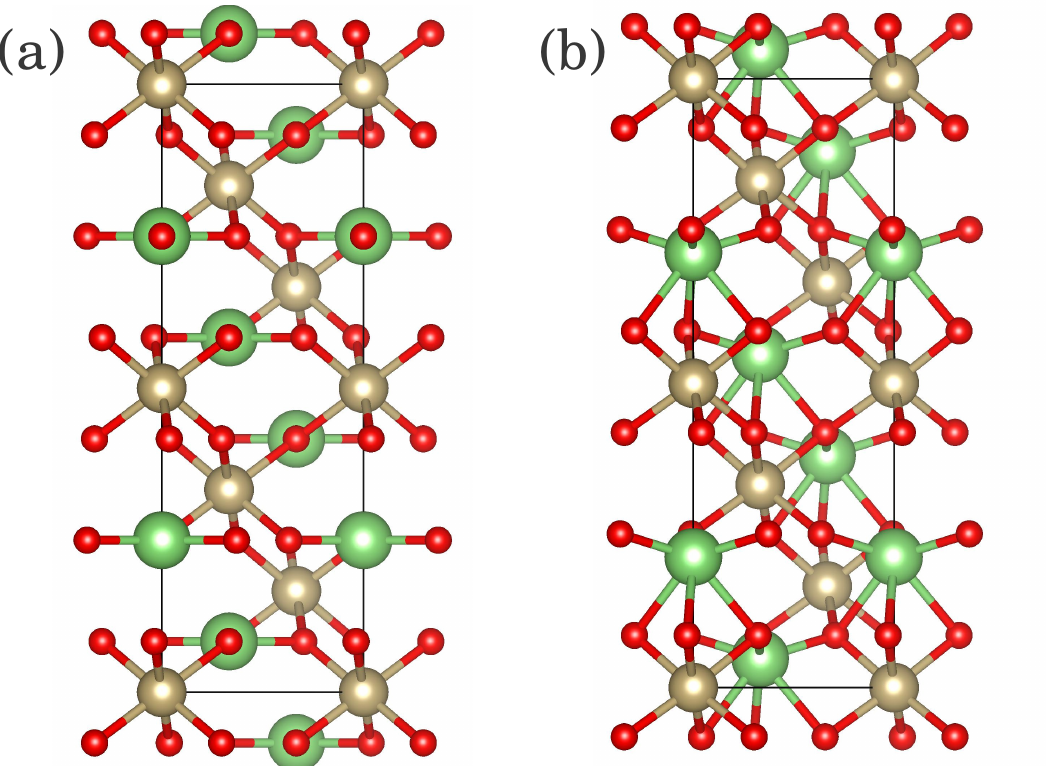}
  \caption{\textbf{Crystal structure of LiOsO$_3$.}  (a) The high symmetry, and (b) the low symmetry phases of LiOsO$_3$. The green, gold and red spheres denote the Li, Os and O atoms, respectively. In the low symmetry phase the Li atoms are displaced from the high symmetry positions resulting in a polar structure.}  \label{structure}
\end{figure}

LiOsO$_3$ undergoes a phase transition from a non-polar $R\overline{3}c$ structure to a polar $R3c$ one (see Fig.~\ref{structure}) at a temperature of 140 K~\cite{shi2013ferroelectric}. This is a continuous transition analogous to the ferroelectric transitions in LiNbO$_3$ and LiTaO$_3$, which occur at significantly higher temperatures. LiOsO$_3$ remains metallic across the transition, thus emerging as a prototypical polar metal. This discovery has led to a wide interest in understanding the properties of LiOsO$_3$. The origin of the polar distortion has been theoretically investigated and has been variously attributed to an A-site cation instability~\cite{xiang2014origin,sim2014first} and an anisotropic unscreened Coulomb interaction~\cite{liu2015metallic}. Further, the role of electronic correlations in this system has also been studied~\cite{giovannetti2014dual,vecchio2016electronic}. On the experimental side, the Raman phonons in LiOsO$_3$ have been measured~\cite{jin2016raman}. Temperature-dependent second harmonic generation experiments have provided evidence for a continuous polar phase transition~\cite{padmanabhan2018linear}. Very recently, a pressure-induced enhancement of the transition temperature has been reported~\cite{paredes2018pressure}.

Over the years, owing to rapid experimental developments, strain has emerged as a very useful tool to engineer and tune the properties of a wide variety of oxide materials~\cite{schlom2007strain,schlom2014elastic}. For instance, enhancement of ferroelectricity in barium titanate~\cite{choi2004enhancement} and room-temperature ferroelectricity in quantum paraelectric strontium titanate~\cite{haeni2004room} have been demonstrated using strain. Tuning the spin-lattice coupling by means of strain has also allowed engineering strong multiferroicity~\cite{lee2010strong}. Charge doping of perovskite oxides, both by chemical and electrostatic means, is an indispensable approach to create new interesting states. Electrostatic doping of insulating oxides, ranging from cuprates to quantum paraelectrics strontium titanate and potassium tantalate, has been used to engineer superconductivity~\cite{mannhart1991electric,ueno2008electric,ueno2011discovery}. Recently, chemical doping has been employed to create polar metal state starting from conventional ferroelectric materials~\cite{fujioka2015ferroelectric,takahashi2017polar,gu2017coexistence}.

\begin{figure}
\includegraphics[scale=0.6]{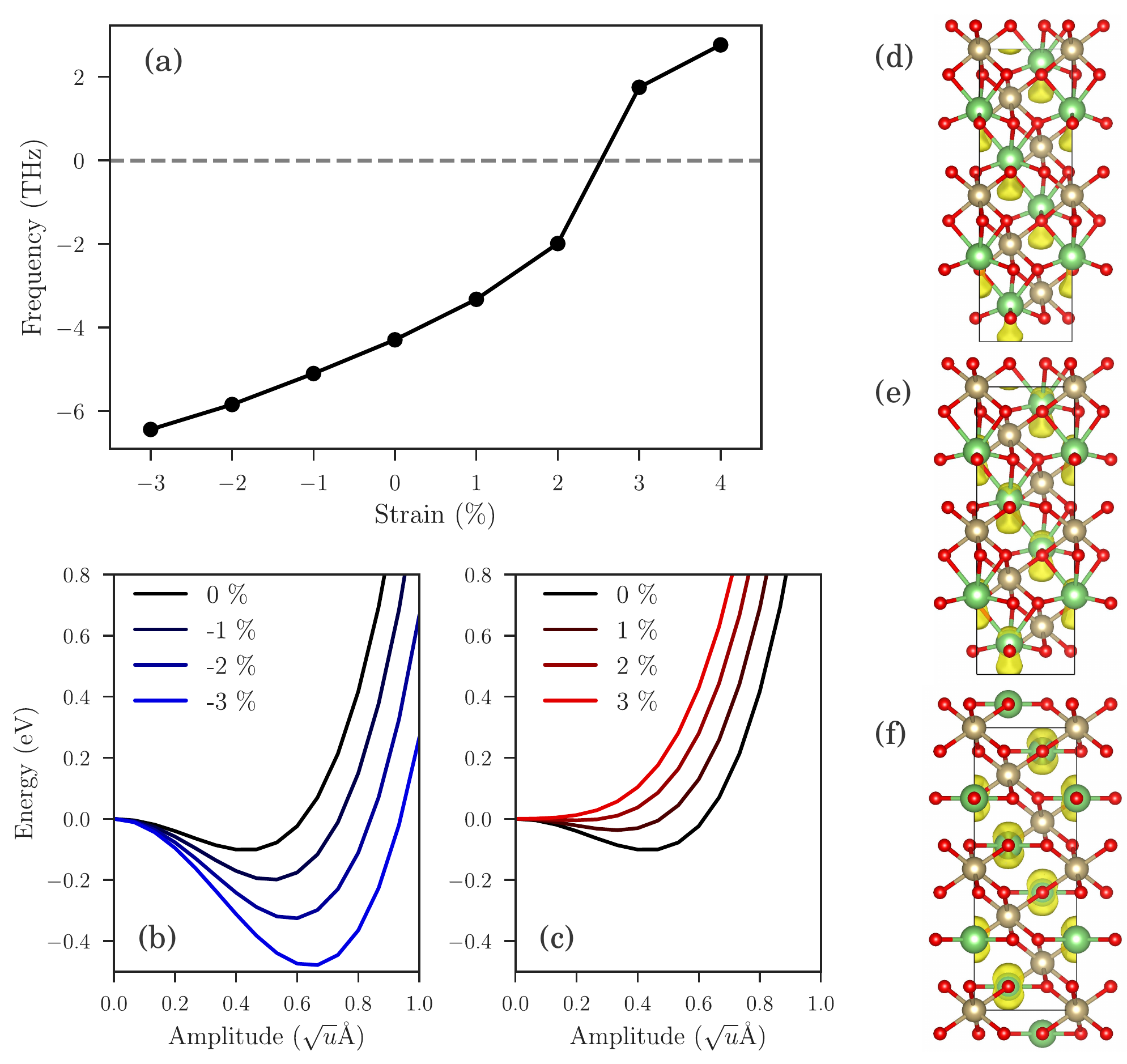}
  \caption{\textbf{Strain control over the polar metal phase.} (a) The frequency of the polar soft mode ($A_{2u}$) with varying biaxial strain. Here negative frequencies denote imaginary modes. Compressive strain makes LiOsO$_3$ more unstable towards the $A_{2u}$ mode distortion. Tensile strain increases the frequency of this mode. A strain-driven quantum phase transition occurs at around 2.5\% tensile strain, where the $A_{2u}$ mode becomes stable. The energy as a function of the polar mode amplitude at different values of (b) compressive, and (c) tensile biaxial strain. Energy gain associated with the polar distortion increases with increasing compressive strain. Increasing tensile strain reduces the depth of the energy well, eventually eliminating the minimum. Here $u$ is the atomic mass unit. Charge density for structures at (d) 3\% compressive strain, (e) zero strain and (f) 3\% tensile strain. The asymmetry of charge density around the Li atoms increases with increasing compressive strain, while for increasing tensile strain the charge density becomes more symmetric, signaling a non-polar phase. An iso-surface value of 0.005 was chosen for all plots.}  \label{strain_frequency_energy}
\end{figure}

In this work, we demonstrate that the polar metal phase in the canonical Anderson-Blount material LiOsO$_3$ can be controlled by biaxial strain. Based on first-principles density functional calculations, we show that a compressive biaxial strain enhances the stability of the polar $R3c$ phase. On the other hand, a tensile biaxial strain favors the centrosymmetric $R\overline{3}c$ structure. Thus, strain emerges as a promising control parameter over polar metallicity in this material. Further, we uncover a strain-driven quantum phase transition at around 2.5\% tensile strain. We highlight intriguing properties that could emerge in the vicinity of this polar to non-polar metal transition. Next, we examine the effect of charge doping on the polar distortions in LiOsO$_3$. By means of electrostatic doping and supercell calculations with O and Li vacancies, we find that screening from additional charge carriers, which are expected to suppress the polar distortions, have only a small effect on them. In contrast to conventional ferroelectrics, where charge doping suppresses ferroelectricity, we find that the polar metal phase in LiOsO$_3$ remains robust against charge doping up to large doping values. \\

\begin{figure*}
\includegraphics[scale=0.55]{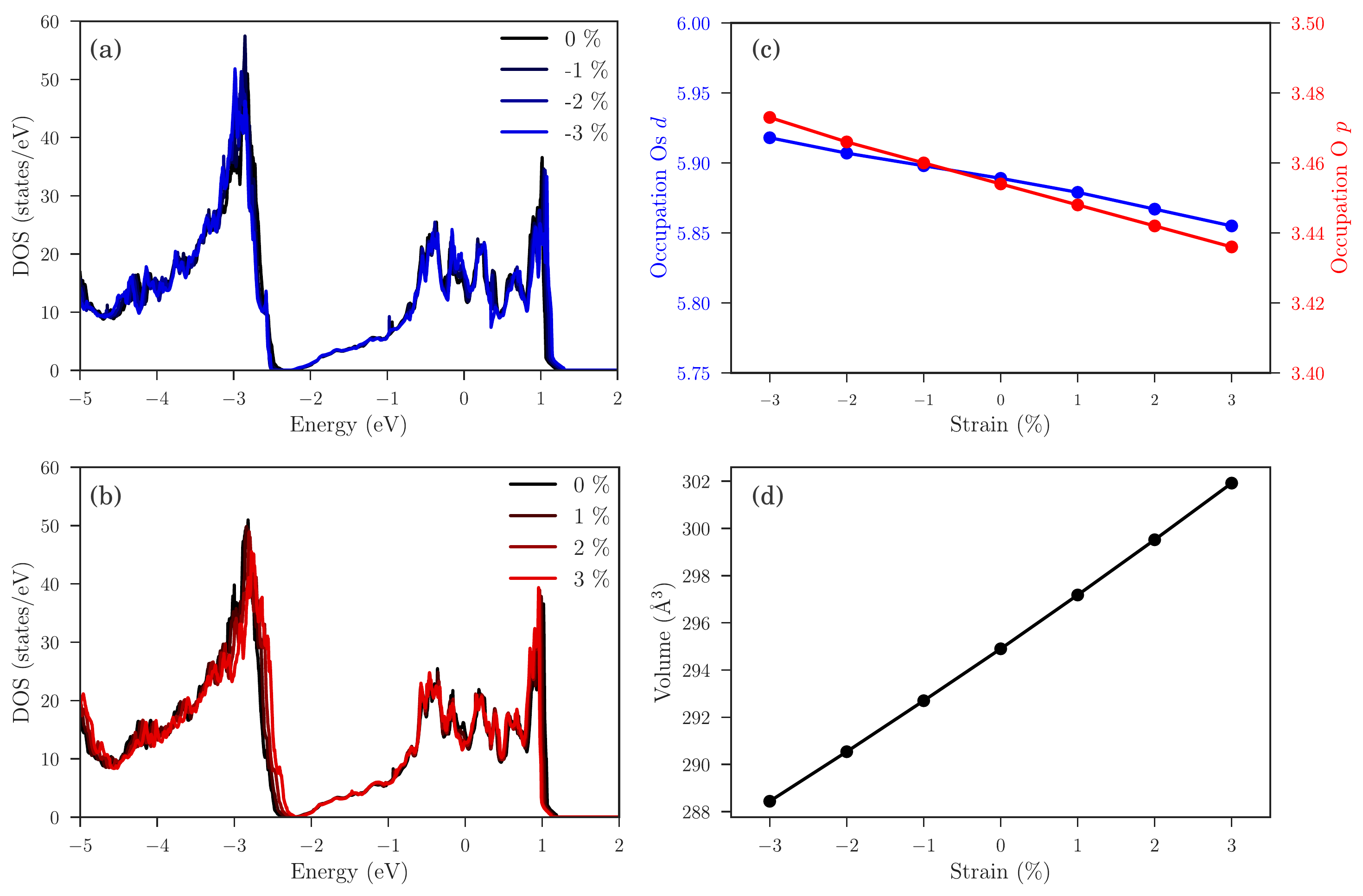}
  \caption{\textbf{Density of states, occupations and volume with varying strain.} Energy dependence of the density of states (DOS) for different values of (a) compressive and (b) tensile biaxial strain. The Fermi level is located at zero energy. The system remains metallic for the strain values considered. (c) Occupation of Os $d$ states (left axis) and O $p$ states (right axis). Both change by less than $\sim$1\% on applying strain. (d) Strain dependence of the unit cell volume. Compressive biaxial strain reduces the volume, while tensile biaxial strain increases the cell volume in a nearly linear manner. Note that the experimentally-measured volume of the low symmetry structure is less than that of the high symmetry one.}  \label{strain_dos_occupation_volume}
\end{figure*}

\section{Methods} 

Density functional theory calculations were carried out using the Vienna Ab Initio Simulation Package (VASP)~\cite{kresse1996efficiency,kresse1996efficient}. We used the PBEsol functional, which is a revised Perdew-Burk-Ernzerhof approximation to the exchange correlation functional, that is particularly suited for the description of lattice properties~\cite{perdew2008restoring}. An energy cutoff of 500 eV was used. Biaxial strain was applied to the pseudo-cubic axes of the conventional 30 atom cell. In this case the Brillouin zone was sampled using an $8 \times 8 \times 4$ $k$-point grid. All structures were relaxed until the forces were less than 0.001 eV/\AA{}. Biaxial strain was imposed by fixing the in-plane lattice parameters, while allowing the out-of-plane lattice parameter and all internal coordinates to relax. For doping under electrostatic condiditons, we added extra charge (holes or electrons) to the primitive 10 atom unit cell. A compensating background charge was incorporated to maintain charge neutrality. We used a $12 \times 12 \times 12$ $k$-point mesh. For supercell calculations, we used a $2 \times 2 \times 2$ supercell comprising of 80 atoms. We removed either O or Li atoms to create structures with vacancies, which were then relaxed till the forces were below the threshold of 0.001 eV/\AA{}. To compute the phonon properties, we used the frozen-phonon approach as implemented in the phonopy code~\cite{togo2015first}.\\

\section{Results}

\subsection{Control by strain}

We begin by presenting the phonon frequencies in the $R\overline{3}c$ high symmetry phase. In the absence of strain we find that the $A_{2u}$ mode, which is the polar soft mode leading to the ferroelectric-like distortion, is unstable. The calculated frequency is -4.29 THz (here negative sign indicates an imaginary frequency), which is in good agreement with previous calculations~\cite{xiang2014origin,sim2014first}. First we consider the case of compressive strain [indicated by negative values in Fig.~\ref{strain_frequency_energy}(a)]. Increasing values of compressive strain lead to the $A_{2u}$ mode becoming more unstable. The frequency reaches -6.44 THz at strain values of -3\%. On the other hand, application of tensile strain [positive values in Fig.~\ref{strain_frequency_energy}(a)] leads to the opposite behavior. The $A_{2u}$ mode becomes more stable, i.e. the frequency becomes less imaginary. At nearly 2.5\% tensile strain we find a strain-driven quantum phase transition from a polar to a non-polar phase, at which point the $A_{2u}$ mode frequency is tuned to zero. Beyond this strain value LiOsO$_3$ remains in the inversion symmetric structure down to zero temperature. To evaluate the structural stability under strain, we have also checked the full phonon spectra for the strained structures. We have checked the phonon dispersions including a Hubbard $U$ term in our DFT calculations. The low frequency spectrum, and in particular the $A_{2u}$ mode frequency, remains practically unchanged.

\begin{figure*}
\includegraphics[scale=0.55]{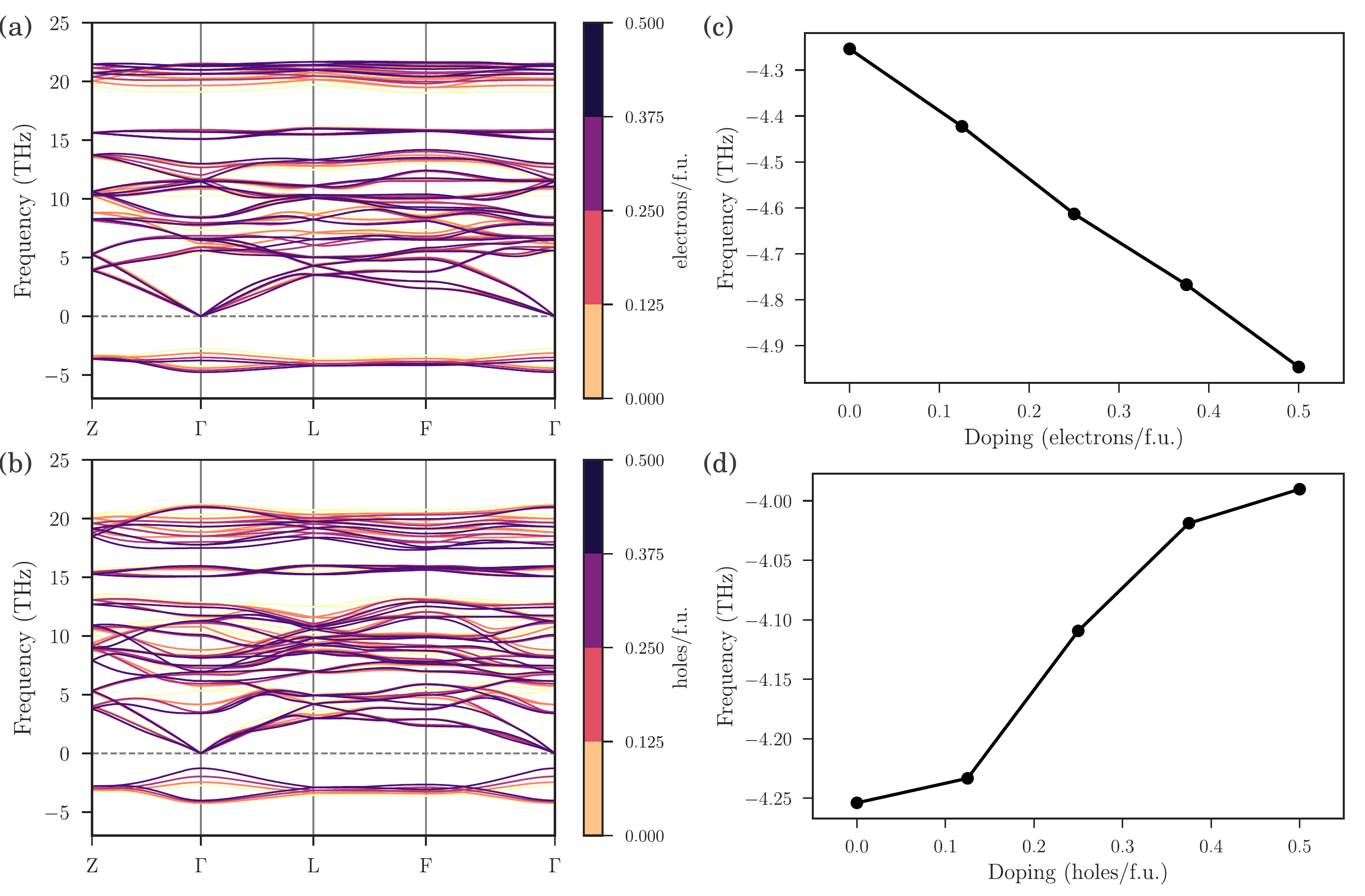}
  \caption{\textbf{Charge doping dependence of the phonons in $R\overline{3}c$ structure.} Phonon spectrum of the high symmetry $R\overline{3}c$ phase with (a) electron doping and (b) hole doping. Note that the negative values indicate imaginary phonon modes. The variation of the soft polar mode frequency at the zone center with (c) electron doping and (d) hole doping. Under both types of charge doping, the polar mode remains unstable.}  \label{charge_doping_phonon_spectrum}
\end{figure*}

Next, we freeze-in the $A_{2u}$ mode in the high symmetry structure while varying the phonon mode amplitude. The energy as a function of the mode amplitude is presented in Fig.~\ref{strain_frequency_energy}(b) and (c) for compressive and tensile strains, respectively. In the unstrained case, the energy gain on freezing-in the $A_{2u}$ mode is nearly 17 meV per formula unit. This value can be substantially enhanced by applying compressive strain and the energy gain reaches 79 meV per formula unit at 3\% compressive strain. This energy difference is proportional to the transition temperature~\cite{wojdel2013controls}. Therefore, we predict a substantial enhancement of the non-polar to polar metal transition temperature under compressive strain. In contrast, under tensile strain we find that the energy gain by freezing-in the $A_{2u}$ mode is progressively reduced. Beyond nearly 2.5\% tensile strain the minimum in energy is lost, signaling the suppression of the structural transition temperature to zero. This tuning with strain, which is a non-thermal control parameter, allows us to tune LiOsO$_3$ through a quantum phase transition. At values of tensile strain higher than the critical value, the high symmetry structure remains more stable and there is no non-polar to polar structural transition with temperature. In recent work, Lu and Rondinelli have demonstrated a polar to non-polar transition in layered oxides by means of both compressive as well as tensile strains, focussing on insulating systems~\cite{lu2016epitaxial}. On the other hand, in the present case we find that only a tensile strain leads to a non-polar to polar transition in a metallic system. To get an insight into the bonding we performed an analysis of the charge density with varying strain values as we show in Fig.~\ref{strain_frequency_energy}(d)-(f). We note that the asymmetry of charge density around the Li atoms increases with increasing compressive strain, while for increasing tensile strain the charge density becomes more symmetric, signaling a non-polar phase.

To check if any electronic transitions accompany our prediction of this strain tuning of the polar metal phase, we compute the density of states at different values of strain, as shown in Fig.~\ref{strain_dos_occupation_volume}(a)-(b). The system remains metallic at all strain values that we considered. In fact, the densities of states are almost identical at different strains. This confirms that the metallic nature of LiOsO$_3$ remains robust under the application of strain. We further calculate the orbital occupations by integrating the charge density over the volume of a sphere, which is centered at the position of the ions. These are shown in Fig.~\ref{strain_dos_occupation_volume}(c) for Os $d$ and O $p$ states. Both change by less than 1\% with the applied biaxial strain. This is again indicative of the fact that in LiOsO$_3$ the electronic structure is only marginally affected by strain.

Using different substrates to grow thin films of LiOsO$_3$ could be a potential way to apply tensile and compressive strains and experimentally pursue our predictions. Different oxide materials could prove to be useful in this regard. For instance, PbO$_2$ and SiO$_2$ could be used as substrates to apply compressive strains of 2\% and 1.7\%, respectively. On the other hand, ternary oxides CdMoO$_4$, LiAlO$_2$, CaWO$_4$ and YAlO$_3$ would yield tensile strain values of 1.9\%, 2.1\%, 2.6\% and 3.5\%, respectively.

A possible reason why a compressive strain favors the low symmetry $R3c$ structure, while a tensile strain promotes the high symmetry $R\overline{3}c$ phase can be the volume effect, which has been very recently put forth by Paredes-Aulestia \textit{et al.}~\cite{paredes2018pressure}. In a nice joint theoretical and experimental work, they showed that application of pressure enhances the structural transition temperature in LiOsO$_3$. The low symmetry $R3c$ structure has a lower volume than the high symmetry structure~\cite{shi2013ferroelectric,paredes2018pressure}. Since, higher pressure favors the structure with lower volume, the transition temperature is enhanced with increasing pressure. To check if this idea is consistent with our findings under biaxial strain, we plot the volume of the cell with the applied biaxial strain in Fig.~\ref{strain_dos_occupation_volume}(d). We note that the relaxed cell volume is indeed smaller under compressive strain, in which case we find that the low symmetry $R3c$ phase has enhanced stability. On the other hand, tensile strain leads to an enhanced volume, as well as an increased stability of the high symmetry $R\overline{3}c$ phase. Such a volume effect is therefore consistent with our findings under biaxial strain, and helps explain the relative stability of the two phases. We have checked that different functionals consistently show that the volume of the polar phase is smaller than that of the non-polar one, consistent with the experimental reports. We note that in some LiNbO$_3$-type materials, such as LiNbO$_3$ and ZnSnO$_3$, the volume of the polar phase is smaller than the non-polar phase. In these systems there are existing predictions, which await experimental confirmation, that a compressive strain should enhance stability of the polar structure~\cite{gu2018cooperative}. This behaviour is in contrast to BaTiO$_3$-type ferroelectrics where a compressive strain reduces the polarization.\\

\subsection{Robustness with doping}

\begin{figure*}
\includegraphics[scale=0.55]{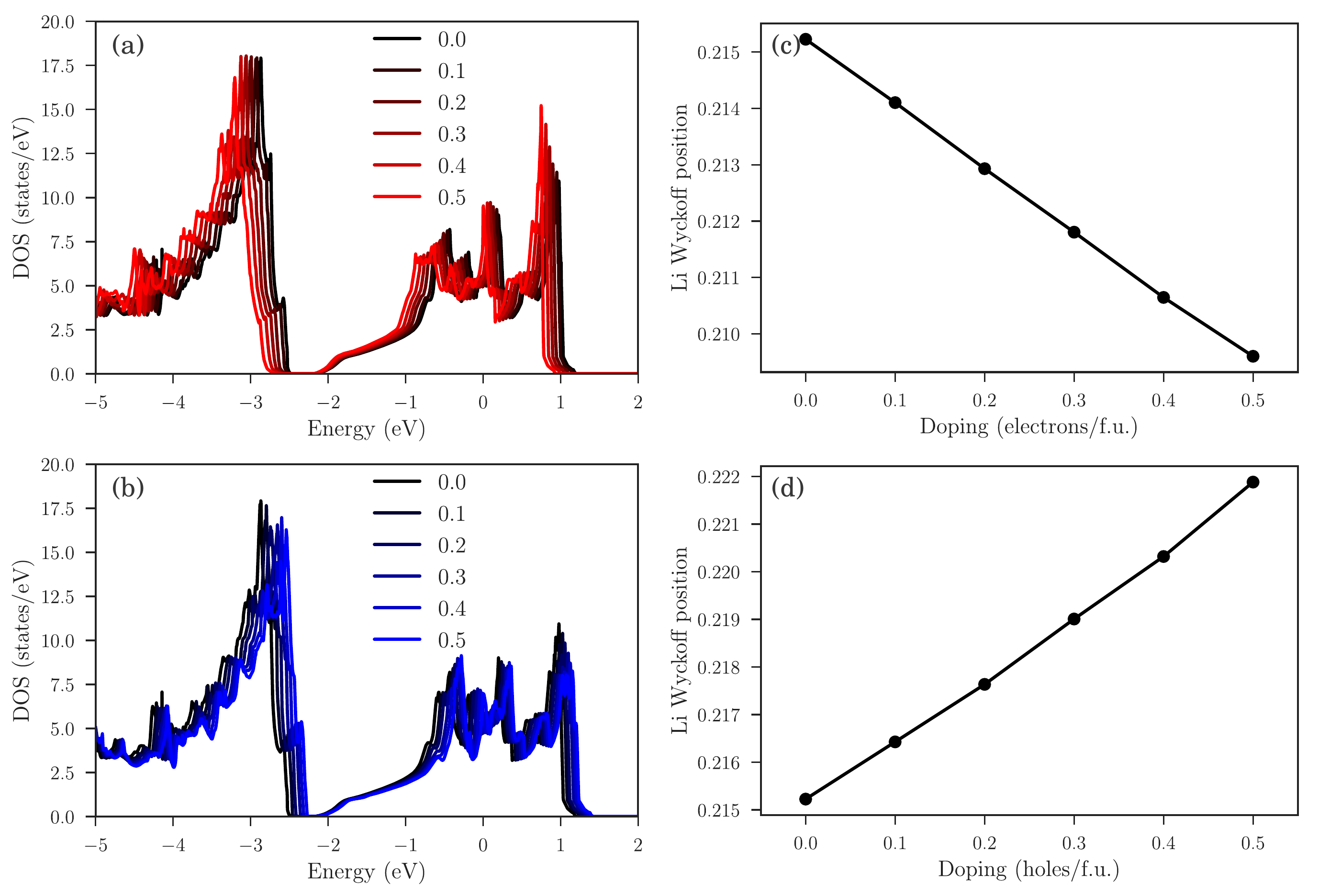}
  \caption{\textbf{Density of states and Li Wyckoff positions with electrostatic doping.} Density of states of the low symmetry $R3c$ phase with (a) electron doping and (b) hole doping. (c) Electron doping and (d) hole doping dependence of Li Wyckoff position in the low symmetry phase. Note that the Li Wyckoff position takes the value of 0.25 in the high symmetry $R\overline{3}c$ structure.}  \label{charge_doping_dos_li_wyckoff}
\end{figure*}

After examining the role of strain in the polar metal phase of LiOsO$_3$, we next move on to study the effect of doping on this phase. Usually one expects addition of charge carriers to screen the dipolar interactions and thereby suppress the polar distortions. To investigate this effect in LiOsO$_3$, we begin by calculating the full phonon spectrum in the high symmetry $R\overline{3}c$ structure under different charge doping conditions. We consider doping values of up to 0.5 electrons or holes per formula unit. The obtained phonon dispersions are presented in Fig.~\ref{charge_doping_phonon_spectrum}(a) and (b), with electron and hole doping, respectively. The obtained phonon spectrum for the undoped case are in good agreement with previously calculated reports~\cite{xiang2014origin,sim2014first}. Most noticably, we find that the instability of the $A_{2u}$ mode, and therefore the polar metal phase, remains robust with both electron and hole doping. At the maximum electron doping value considered (0.5 electrons per formula unit), the $A_{2u}$ mode frequency at the zone center becomes more imaginary by nearly 13\%. On the other hand, a hole doping of 0.5 per formula unit results in the $A_{2u}$ mode frequency becoming less imaginary by approximately 6\%. In both cases the polar mode remains unstable up to a substantial doping value of 0.5 additional carriers per formula unit. We note that this is in sharp contrast to the case of prototypical ferroelectric BaTiO$_3$, where a doping of 0.1 electrons per formula unit has been shown to be sufficient to completely eliminate the polar mode instability~\cite{wang2012ferroelectric}.

\begin{figure*}
\includegraphics[scale=0.55]{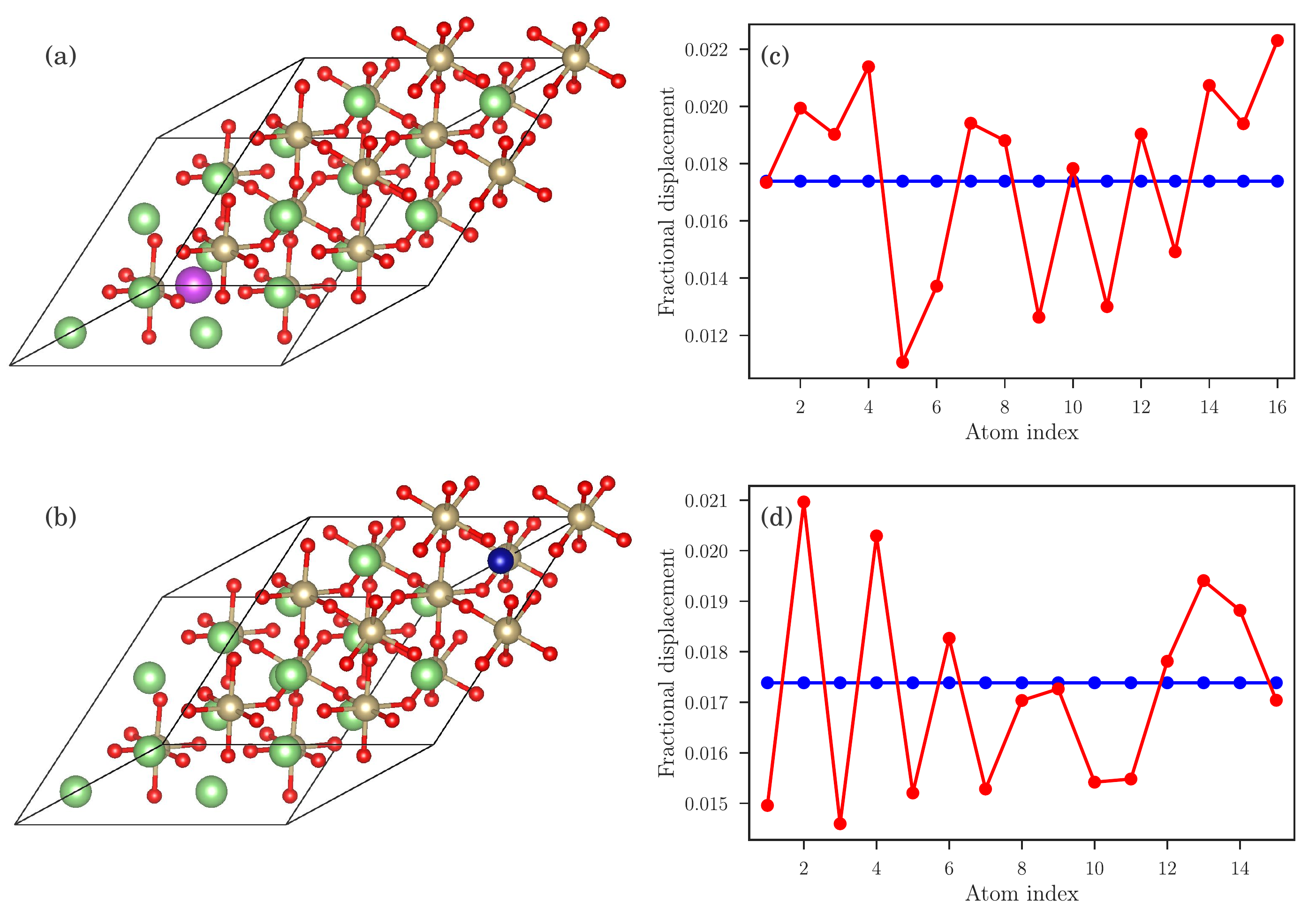}
  \caption{\textbf{Chemical doping with oxygen and lithium vacancies in $R3c$ structure.} (a) Electron doping induced by an O vacancy. The purple sphere indicates the position of the O vacancy. (b) Hole doping by a Li vacancy. Here the blue sphere denotes the position of the Li vacancy. Fractional displacements of the Li atoms from the high symmetry positions in the case of (c) O vacancy and (d) Li vacancy. The blue lines in both plots correspond to the reference case of no vacancy, while the red lines are the displacements in the presence of the vacancy.}  \label{charge_doping_vacancy_displacement}
\end{figure*}

To further explore this robustness of the polar metal phase under charge doping, we next consider the low symmetry polar $R3c$ structure. We relaxed the structure at the different values of charge doping up to 0.5 carriers per formula unit. The densities of states for electron and hole doping are shown in Fig.~\ref{charge_doping_dos_li_wyckoff}(a) and (b), respectively. Under the adopted electrostatic doping conditions, the doping essentially leads to a rigid shift of the Fermi level to accommodate the additional carriers. Atomic relaxations at different doping values have a negligible effect on other features of the densities of states. Since, the Li atoms are the lightest and therefore show the largest contribution to the polar distortion, we extract their Wyckoff positions for each of relaxed structures at different doping values. Note that in the high symmetry $R\overline{3}c$ structure this Wyckoff position takes the value 0.25 (corresponding to zero polar distortion). As shown in Fig.~\ref{charge_doping_dos_li_wyckoff} (c) and (d), we find that both with electron as well as hole doping Li Wyckoff position remains close the undoped value of 0.215, and therefore the structure remains polar. With electron doping of 0.5 electrons per formula unit the Li Wyckoff position decreases slightly to 0.210, while a hole doping of 0.5 results in a Li Wyckoff position of approximately 0.222. Overall this is a variation of less than 3\% from the undoped value. We can conclude that the polar displacements in LiOsO$_3$ are remarkably robust with doping. This robustness to both electron and hole doping can be understood based on the decoupling of the itinerant electrons (Os $d$ states) from the polar phonons (composed mainly of Li displacements) which drive the polar metal transition as has been noted in the pioneering work of Anderson and Blount~\cite{anderson1965symmetry} and more recently by Puggioni and Rondinelli~\cite{puggioni2014designing}. Upon both electron as well as hole doping, itinerant electrons lie in Os $d$ states, which allows decoupling from the Li displacements driving the polar distortion. This may be the underlying reason for such a robustness of the polar metal phase in LiOsO$_3$.

Until now we have simulated the effect of doping by adding electric charge (along with a compensating background) to the simulation cell. To further confirm our conclusions regarding the robustness of the polar metal phase, we carried out supercell calculations with O and Li vacancies. O vacancies lead to an electron doping, while Li vacancies result in a hole doping of the system. We constructed $2\times 2\times 2$ supercells of the polar $R3c$ structure comprising of 80 atoms in total. Then we removed either one O atom or one Li atom and the resulting structures are presented in Fig.~\ref{charge_doping_vacancy_displacement}(a) and (b). This yields an O vacancy concentration of approximately 2\% (corresponding to 0.125 additional carriers per formula unit) and a Li vacancy concentration of nearly 12.5\% (corresponding to 0.0625 additional carriers per formula unit). We relaxed the obtained structures and extracted the fractional displacements of the Li atoms from their high symmetry positions. For reference we also extracted these for the case of no vacancy, i.e., no doping. The reference displacements are shown by blue lines in Fig.~\ref{charge_doping_vacancy_displacement}(c) and (d). Note that in the case of no vacancy all Li atoms are equivalent and therefore have the same fractional displacement with reference to the high symmetry structure. The displacements of the Li atoms with O or Li vacancies are shown by red lines. First consequence of creating vacancies is that there is now a structural symmetry breaking and different Li atoms are no longer equivalent. This inequivalence leads to different fractional displacements for different Li atoms. In the case of O vacancy, for a majority of the Li atoms the displacement compared to the pristine case increases marginally by around 8\%. In the case of Li vacancy, a majority number of Li atoms show a reduction in the polar displacement, which does not exceed approximately 14\%. In both cases, all Li atoms display polar displacements. In summary, we note that in both cases of electron (O vacancy) and hole (Li vacancy) chemical doping, the polar displacements remain robust in agreement with our conclusions from electrostatic doping. Importantly, with an eye on experiments where samples inevitably have vacancies, such charge doping defects should not suppress the polar metal phase in LiOsO$_3$.

Here we have focussed on analysing the effect of oxygen and lithium vacancies on the charge doping. More generally, vacancies can introduce a variety of other important effects. For instance, vacancies can act as trapping centres for electrons or holes, which can then be effectively tuned using strain~\cite{lopez2015magnetism}. Furthermore, vacancies can introduce magnetic moments and potentially lead to long-range magnetic order~\cite{cheng2016manipulation}. Vacancies can also have profound effects on trasport properties of materials. For example, O vacancies have been suggested to lead to $n$-type conductivity in zinc oxide~\cite{liu2016oxygen}. It would be worthwhile exploring these and other intriguing effects associated with vacancies in more detail.

\section{Discussion} 

Our work highlights strain as a promising control parameter over the polar metal phase in LiOsO$_3$, as well as the robustness of this phase against charge doping. The tuning to a critical state between a polar and a non-polar metal under tensile strain is particularly exciting. In proximity to criticality, materials tend to be susceptible to fluctuations and one expects intriguing phases to emerge~\cite{narayan2019multiferroic}. Close to the critical value of strain LiOsO$_3$ can be a promising ground to explore parity-breaking gyrotropic and multipolar orders~\cite{fu2015parity}. Various properties, such as susceptibilty and transport observables, show anomalous behavior near critical points. Indeed, such an anomalous enhancement of thermopower has been reported in Mo$_x$Nb$_{1-x}$Te$_2$ system tuned to the polar to non-polar phase boundary~\cite{sakai2016critical}. It would be exciting to perform similar experiments in LiOsO$_3$ close to the critical strain value predicted by us. Recently, topological Dirac points and nodal rings have been revealed in LiOsO$_3$ by means of density functional calculations~\cite{yu2018nonsymmorphic}. Our results also suggest an experimentally-readily-accessible handle to control these topological semimetal features by using strain and tuning the Fermi level to the topological crossing points by charge doping. Strain may also be used to control the properties of ultrathin LiOsO$_3$, which has recently been proposed to exhibit asymmetric hysteresis~\cite{lu2019ferroelectricity}. Interplay of charge doping and quantum confinement effects in ultrathin LiOsO$_3$ could be worth investigating in future. It would be worthwhile to explore the role of strain in magnetically driven polar metals, one example of which has been very recently synthesized~\cite{princep2019magnetically}. It would also be interesting to investigate whether the magnetic polar metals are robust to doping and how doping affects the magnetic order in these systems.  \\

\section{Summary} 

By using first-principles density functional calculations, we demonstrated that the polar metal phase in the prototypical Anderson-Blount material LiOsO$_3$ can be effectively controlled by biaxial strain. Compressive strain allows increasing the stability of the polar metal phase, while a tensile strain favors the non-polar structure. We showed that under experimentally-accessible values of tensile strain, a quantum phase transition from a polar to a non-polar metal can be achieved. Next, we examined the effect of electron and hole doping on the polar metal phase. By means of electrostatic doping as well as supercell calculations with vacancies, we found that screening from additional charge carriers, expected to suppress the polar distortions have only a small effect on them. Rather remarkably, and in contrast to conventional ferroelectrics, polar metal phase in LiOsO$_3$ remains robust against charge doping up to a large doping value of 0.5 additional charge carriers per formula unit. In conclusion, our hope is that these findings stimulate further theoretical and experimental explorations into control over polar metals and their intriguing properties. \\

\section*{Acknowledgments} I would like to thank Nicola Spaldin and Naga Phani Aetukuri for helpful suggestions, and Alexey Soluyanov and Dominik Gresch for useful discussions on topological aspects of lithium osmate. I acknowledge support from Indian Institute of Science and ETH Zurich. Computational resources were provided by ETH Zurich (Euler cluster).

\bibliography{references}

\end{document}